# Integrated high quality factor lithium niobate microdisk resonators

**Cheng Wang,[1,*] Michael J. Burek[1], Zin Lin[1], Haig A. Atikian[1], Vivek Venkataraman[1], I-Chun Huang[1], Peter Stark[1] and Marko Lončar[1,*]**

[1] *School of Engineering and Applied Sciences, Harvard University, 29 Oxford Street, Cambridge, MA 02138, USA*
[*]*chengwang@seas.harvard.edu*
[*]*loncar@seas.harvard.edu*

**Abstract:** Lithium Niobate (LN) is an important nonlinear optical material. Here we demonstrate LN microdisk resonators that feature optical quality factor ~ $10^5$, realized using robust and scalable fabrication techniques, that operate over a wide wavelength range spanning visible and near infrared. Using our resonators, and leveraging LN's large second order optical nonlinearity, we demonstrate on-chip second harmonic generation with a conversion efficiency of 0.109 $W^{-1}$.

### References and links

---

## 1. Introduction

Single-crystal lithium niobate ($LiNbO_3$, LN), with its high second order susceptibilities ($\chi^{(2)}$), wide transmission window (400 nm to 5 μm, with an $OH^-$ absorption peak at 2.87 μm), relatively high ordinary and extraordinary refractive indices ($n_o$ = 2.21, $n_e$ = 2.14 at 1550 nm, respectively) [1], as well as large piezoelectricity, is an excellent nonlinear optical material for communication science and technology. Specifically, its $\chi^{(2)}$ has the largest component in diagonal direction ($\chi^{(2)}_{zzz}$) and is high for both second harmonic generation (SHG, 41.7 pm/V) and electro-optic (EO) modulation (30.9 pm/V). Conventional LN components, including EO modulators or periodically-poled (PPLN) wavelength conversion devices, are realized in either proton-exchanged or titanium-diffused LN waveguides [2-4], which possess very low index contrast ($\Delta n < 0.1$) between core and cladding. As a result, the poor light confinement requires waveguide dimensions to be many wavelengths in both thickness and width [5-7]. This not only limits the integration capability, but more importantly results in large power requirements for nonlinear optical processes. While researchers have made great efforts to improve these platforms (e.g. PPLN ridge waveguides [8]), it is still difficult to implement on-chip micro-cavities to enhance nonlinear optical processes as those demonstrated in other materials [9-15]. An integrated LN nanophotonic platform, which combines the unique material properties of LN with the superior light confinement in wavelength scale optical waveguides and cavities, could overcome these limitations and enable efficient, low-cost and highly-integrated EO modulators and wavelength conversion systems.

Owing to the recent development of LN thin film fabrication techniques via crystal ion slicing [16, 17], nanophotonic elements including microring resonators [18], photonic crystal cavities [19] and microdisk resonators [20, 21] have already been realized and used to demonstrate on-chip EO modulation and SHG. However, nanostructuring of LN, in particular its dry etching, is challenging and usually results in rough sidewalls, large scattering losses and fairly low optical quality (Q) factors (4,000 in Ref. [18]). In 2012, Wang et al. proposed that, by heating the microdisks close to its melting temperature after device fabrication, local melting on the disk edge could smoothen the surface through surface tension reshape, and the Q factors are subsequently increased to 26,000 [20]. More recently, Q factors as high as 250,000 were obtained using femtosecond laser ablation followed by two-step focused ion beam (FIB) milling [21]. While these results are promising, they utilize non-standard fabrication techniques that cannot be easily scaled up. Alternative hetero-structure approaches that circumvent the etching of LN have also been explored [22, 23], but they suffer from reduced overlap between optical fields and LN's nonlinearity.

Herein, we demonstrate high-Q LN optical microdisk resonators fabricated by simple, robust and standard nanofabrication techniques, which require no post-fabrication processing and material treatment. These resonators operate at both visible and near infrared (NIR) wavelengths, featuring Q-factors as high as $1.02 \times 10^5$. A commercially available LN on insulator (LNOI) substrate, prepared via smart-cut techniques, was used to realize our resonators, which were subsequently fabricated by electron-beam lithography (EBL) and an optimized argon plasma etching process, which yield smooth sidewalls and result in high-Qs. As a proof of concept, we demonstrate on-chip SHG in our devices, with an internal conversion efficiency of 0.109 $W^{-1}$. Further improvements, such as development of dispersion engineered ring resonators with the appropriate phase matching condition, would enhance the nonlinear effects by at least two more orders of magnitude [24].

## 2. Fabrication of LN microdisk resonators

The fabrication procedure used to realize our devices is summarized in Fig. 1 (a). LNOI substrates (by NANOLN) with a 400 nm thick LN device layer on top of 1 μm thick buried silicon dioxide layer, further supported on a single-crystal LN substrate, were used. Patterning of microdisk structures started with deposition of a 15 nm titanium layer on the LNOI substrate by electron beam evaporation. The titanium layer served two purposes: to promote resist adhesion and to provide conduction during subsequent EBL. Hydrogen silsesquioxane (HSQ) based negative-tone electron-beam resist (FOX®-16 by Dow Corning) was spin-coated on top of the titanium-coated LNOI substrate and patterned using EBL (Elionix ELS-F125). Resulting HSQ patterns were approximately 600 nm in height and used as an etch mask to define the LN microdisks. A second blanket electron beam exposure of the resist after development was used to increase the mask hardness, thus enhancing the etching selectivity. LN was then dry etched with argon plasma ($Ar^+$) in an electron-cyclotron resonance (ECR) reactive ion etching (RIE) system (NEXX Cirrus 150). A relatively high RF power (200 W) was used to enable DC bias of ~ 180 V that accelerates $Ar^+$ ions towards the substrate, resulting in physical etching of the LN device layer. A typical etch rate is estimated to be ~ 30 nm/min. Approximately 300 nm of LN could be etched using this approach (leaving ~ 100 nm thick unetched LN layer behind), before significant mask erosion, which could result in pattern distortion and increased surface roughness. Following plasma etching, the remaining HSQ was removed by 2 min of wet etching in 5:1 buffered oxide etch (BOE). While the resulting ridge-like microdisks already supported optical modes, they were hard to be characterized using a fiber taper. To fabricate suspended LN microdisks on top of oxide pedestals, the remaining 100 nm thick LN layer was removed by an additional plasma etching step using the same ECR-RIE condition as before, but this time without any mask. Finally, the devices were undercut in 5:1 BOE (for 5 min) resulting in free-standing microdisks supported by a silica pillar.

Representative scanning electron microscope (SEM) images of a final LN optical resonator, with approximately 300 nm thickness, supported by 1 μm thick silica pedestal, are displayed in Fig. 1 (b). Since the argon plasmas do not bombard the surface perpendicularly, sidewalls in our devices usually make an angle between 35° and 40° with respect to substrate normal direction. Nonetheless, the etched sidewalls are quite smooth with few noticeable rough sites under SEM inspection. Together with the fact that the top and bottom surfaces of our LN device layer have been polished to a surface roughness less than 1 nm, scattering loss is expected to be low, resulting in high optical Q-factors discussed next.

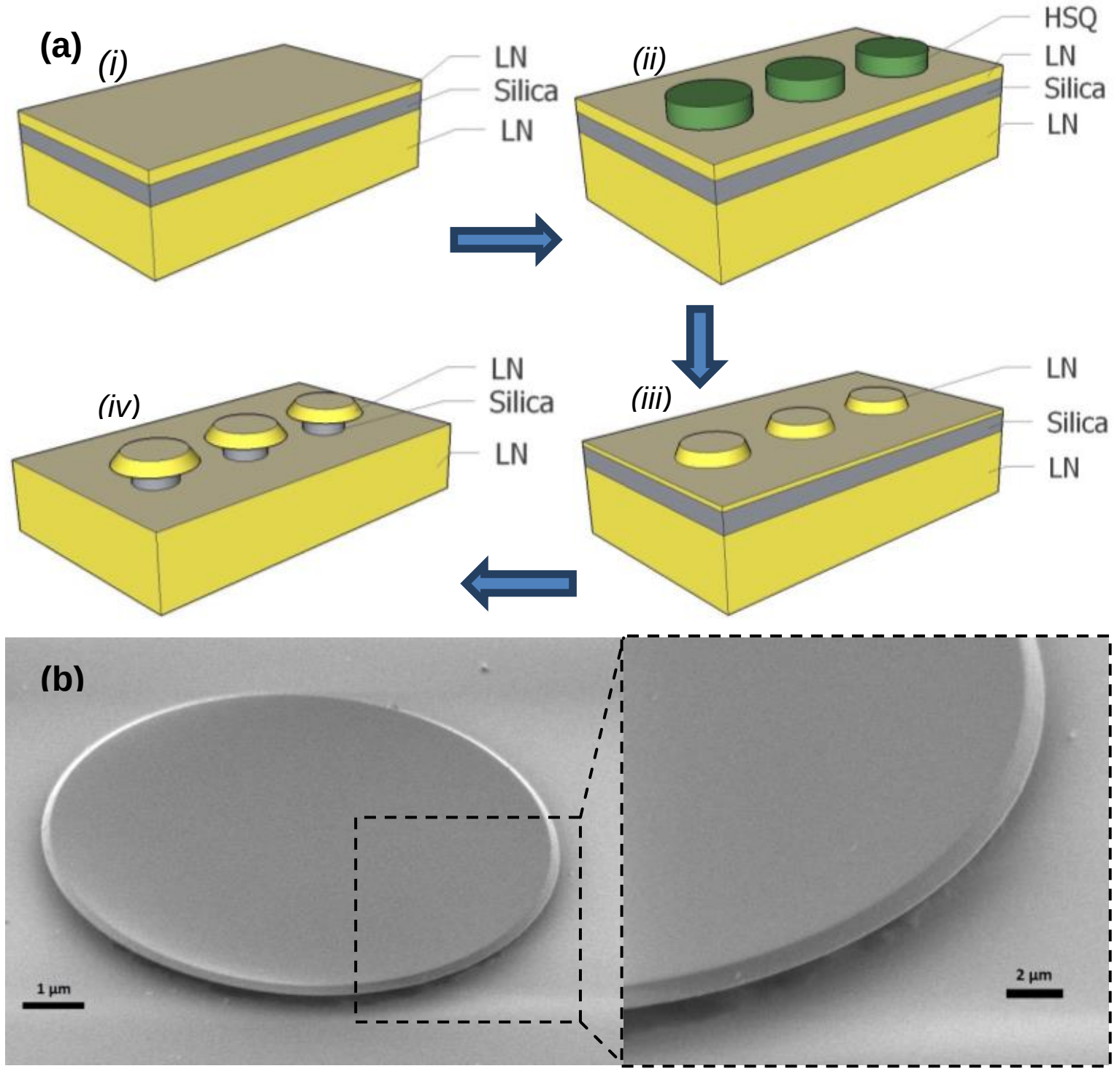

Fig. 1. Fabricated LN microdisk resonators. (a) Fabrication procedure: (i) a LNOI substrate with a 400 nm LN thin film on top of 1 μm silica sacrificial layer, obtained from NANOLN, was (ii) spin-coated with HSQ. EBL is performed to pattern the HSQ resist. (iii) Argon plasma etching was then used to transfer the etch mask pattern into LN, by removing ~ 300nm of LN. HF wet etching was used to remove HSQ residue, followed by (iv) an additional argon plasma etch to globally remove the remaining ~ 100nm of LN. After this step, the LN microdisk devices were defined with a final thickness of ~ 300 nm. Finally, the microdisks were undercut using HF wet etching to form suspended structures. (b) Representative SEM image of a suspended 28 μm diameter LN microdisk supported by a silica pedestal on top of a LN substrate. A magnified view of the microdisk edge reveals smooth sidewalls. All SEM images are taken at a 45° angle.

## 3. Experimental results and discussions

### *3.1 Transmission spectra in telecom wavelengths*

LN microdisks were initially characterized in the telecom wavelength range, with a representative transmission spectrum collected from a 28 μm diameter disk shown in Fig. 2 (a). A silica fiber taper [25-27] was used as a local evanescent probe to couple light into and out of our microdisk resonators. Fiber tapers were produced by flame anneal and pulling from commercial single mode telecom fibers (SMF-28, Dow Corning), resulting in a final diameter of ~ 1 μm. The pulled fiber taper was subsequently mounted in a U-shape and spliced into an optical set-up. The self-tension of the taper region makes it possible to be brought into close proximity to the desired device via motorized stages, as is shown in the inset of Fig. 2 (a). Light from the tunable lasers (Santec TSL-510, tuning range from 1480 to 1680 nm) were sent into the tapered fiber after an inline fiber polarizer, and collected by a high gain InGaAs detector (EO Systems, IGA1.9-010-H). From the spectrum, three different sets of high Q-factor resonant modes are identified and displayed with different colors, each with similar free spectral range (FSR) near 13 nm. Numerical simulations from a finite element mode solver (COMSOL) indicate that these resonances correspond to transverse electric (TE)-like modes with the lowest three radial orders, all with negligible radiation loss. Coupling strength to each mode can vary depending on the fiber position, resulting in different transmission dips for different modes. According to our modeling results, transverse magnetic (TM)-like modes are not supported in our devices since the LN device layer is too thin. Fig. 2 (b-g) show high resolution transmission spectra corresponding to the three different radial mode orders, together with simulated cross-sectional mode profiles. Lorentzian fits to the resonances yield Q-factor estimates for the first, second and third order radial modes of $1.02 \times 10^5$, $8.8 \times 10^4$ and $4.4 \times 10^4$ respectively. The other distinguishable modes with lower Q-factors correspond to even higher radial orders. A large number of resonator modes from many devices were characterized, with different fiber-positions, and we find that nearly all devices support modes featuring Q-factors exceeding $5 \times 10^4$.

### *3.2 Transmission spectra in visible wavelengths*

LN microdisk resonator performance was further characterized in the visible and near-infrared (NIR) wavelength range ($\lambda < 1$ μm). To ensure relatively high modal overlap between the fiber and the microdisk modes for lower wavelengths, fiber tapers with a diameter of ~ 500 to 700 nm are needed. This time the fiber tapers were manufactured by a two-step hydrofluoric acid (HF) wet etching method [28], including ~ 30 minutes of concentrated HF etching, followed by ~ 30 to 45 minutes 5:1 BOE etching. A thin layer of o-xylene was covered on top of the HF solution in order to promote taper formation via the oil/water interface meniscus. Light from a super-continuum source (EXW-4, NKT Photonics) was delivered to the resonators using fiber taper coupling, and the transmission spectrum was collected with an optical spectrum analyzer (OSA, HP 70950B, minimum resolution bandwidth of 0.08 nm). Fig. 3 (a) displays the transmission spectrum of a 30 μm diameter microdisk resonator in the 770 to 780 nm wavelength range. As expected, the microdisk supports large numbers of TE and TM polarized optical modes, and within the FSR of the fundamental TE mode (~ 6 nm) as many as 16 different modes can be identified. The high density of optical modes at these wavelengths is desirable for SHG experiments (discussed next). We note that the line widths of most of the resonances in Fig. 3 (a) are limited by the spectral resolution of our OSA, thus giving only a lower bound of their Q-factors. To obtain better estimates of the Q-factor values, we further characterized our device Q-factors at visible wavelengths by a tunable red laser (New Focus Velocity TLB 6304 laser, coarse tuning range of 634.8 to 638.9 nm and fine tuning range of 70 pm) and a visible band photodetector (New Focus 1801). Fig. 3 (b) shows the collected transmission spectra near 637 nm from the same microdisk resonator as in Fig. 3 (a). The fine tuning mode of the laser was used to accurately measure the Q-factors of cavity modes supported by the LN microdisk resonators within the ~ 4 nm coarse laser tuning range. High resolution spectra of two selected resonance dips with relatively high Q's are shown in Fig. 3 (c-d). The spectra follow Fano-like shapes, likely resulting from interference between light coupling into the microdisk and light reflected from the contact region between the fiber taper and the microdisk [29], or from avoided crossing with other modes that are close in resonance wavelengths. The fine tuning resonance spectra were fitted into a Fano profile [30] , with the relation:

$$F(\lambda) = a_1 - a_2 \cdot \frac{\left(q + \frac{2(\lambda - \lambda_0)}{\gamma}\right)^2}{1 + \left(\frac{2(\lambda - \lambda_0)}{\gamma}\right)^2} \tag{1}$$

where $\gamma$ is the linewidth of the resonance, $\lambda_0$ is the resonance wavelength, q is the Fano parameter related to the asymmetric line shape, $a_1$ and $a_2$ show the baseline and dip depth respectively. Q-factor can be calculated as $\lambda_0/\gamma$.

The Q-factors of the two resonances are thus estimated to be $4.6 \times 10^4$ and $2.6 \times 10^4$ respectively. The highest measured Q-factor in the visible wavelengths is nearly half of that in the telecom wavelengths, likely due to several reasons: (1) resonant modes with higher Q-factors may exist outside our laser range; (2) the much higher mode density at visible wavelengths gives an additional loss channel into higher-order low-Q modes; (3) surface roughness becomes more comparable to the optical wavelengths, inducing more scattering loss.

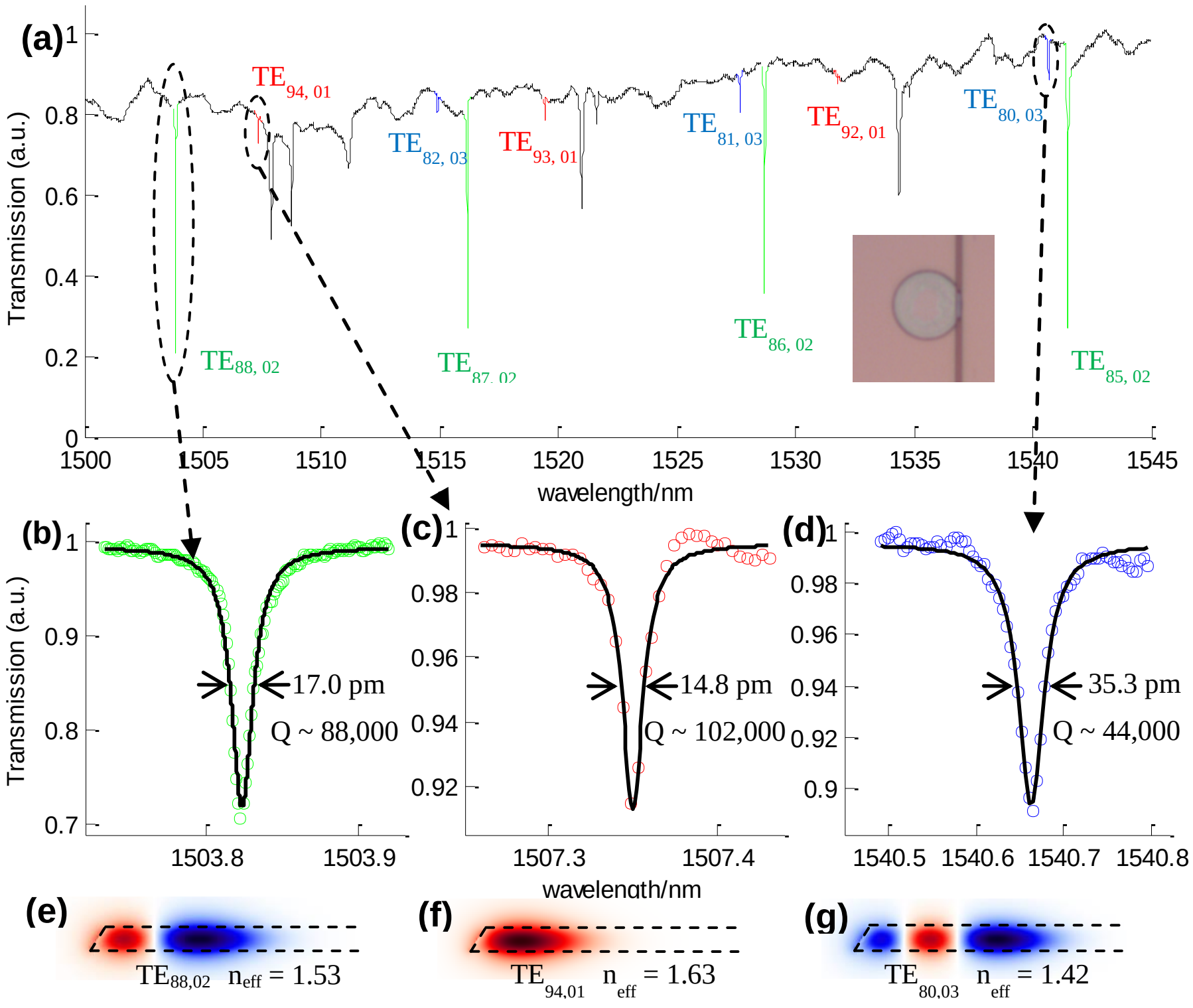

Fig. 2. Resonance spectra and simulation results in telecom wavelengths. (a) A representative transmission spectrum collected from a 28 μm diameter disk revealing several sets of resonances, indicated by color coding. Each resonant dip is labeled as $TE_{mn}$, where m, n are the azimuthal and radial mode numbers respectively. Inset shows the optical micrograph of tapered fiber coupling on top of the microdisk resonator. (b-d) High-resolution views of representative resonance dips for each radial mode and their corresponding Lorentzian fits (black curves), indicating quality factors of $1.02 \times 10^5$, $8.8 \times 10^4$ and $4.4 \times 10^4$ for 1st (c), 2nd (b) and 3rd (d) order radial modes respectively. (e-g) Finite element mode simulation of the three radial order modes shown in (b-d), indicating TE-like modes. Electric fields along radial direction ($E_r$) are shown, labeled with mode numbers and effective indices. Dashed lines show the outlines of the device cross section.

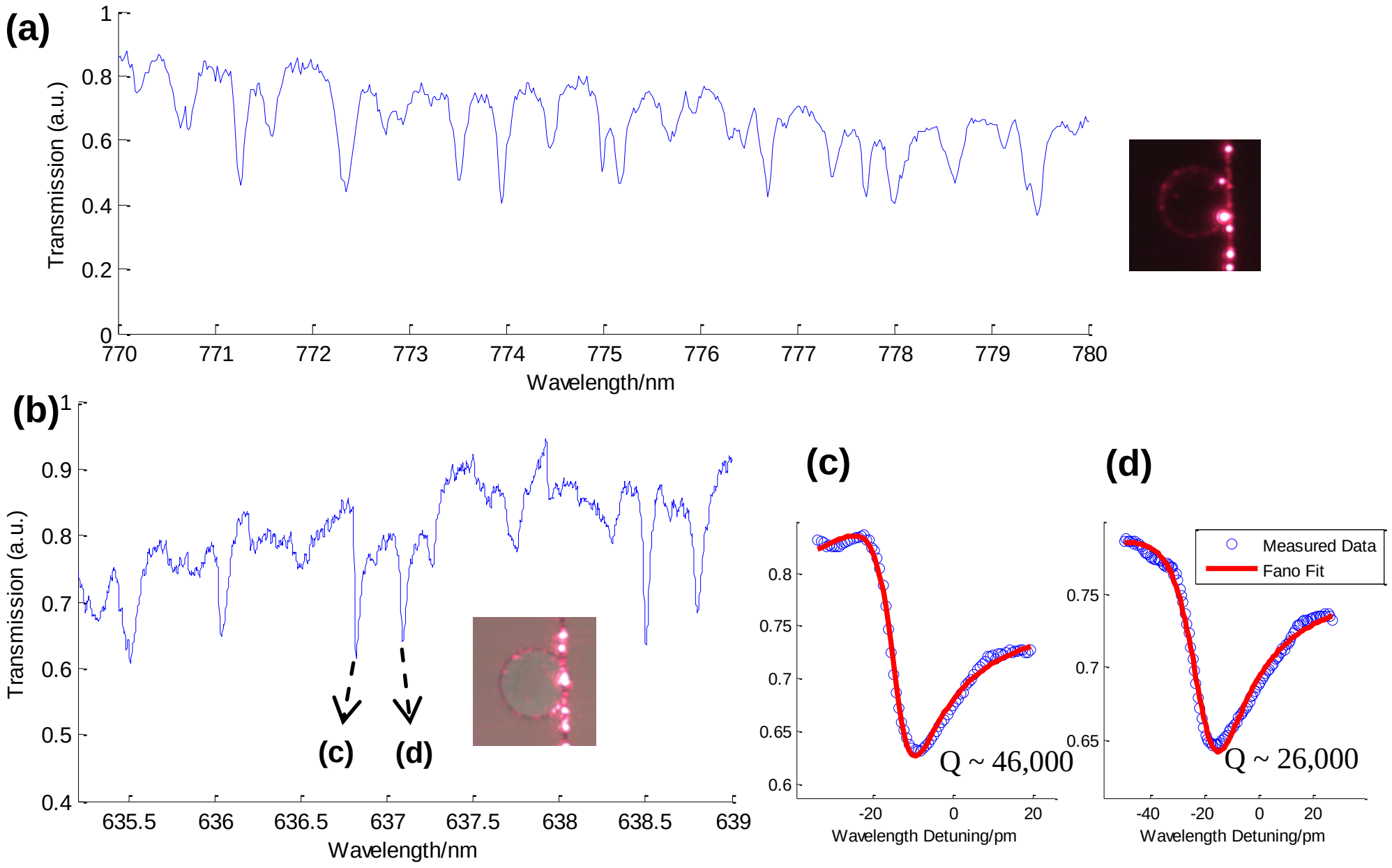

Fig. 3. Resonance spectra in visible wavelengths. (a) Transmission spectrum at second harmonic wavelengths. Inset: optical micrograph showing broad band red light (600 nm – 800 nm supercontinuum) coupling into the microdisk resonator. (b) Transmission spectrum ~ 637 nm obtained from a tunable diode laser. Inset: optical micrograph of red light coupling into the resonator. (c-d) High-resolution spectra of two representative resonance dips near 636.8 nm (c) and 637.1 nm (d), taken in laser wavelength fine tuning mode, and their Fano fits, indicating quality factors of $4.6 \times 10^4$ and $2.6 \times 10^4$.

### *3.3 On-chip second harmonic generation*

To take advantage of the nonlinear optical properties of the single crystal LN microdisk resonator shown in Fig. 3 (a), we tuned a telecom wavelength pump laser into one of its resonances near 1546 nm and monitored the spectrum of the transmitted light using a spectrometer (HORIBA iHR550). The spectrometer data clearly show the presence of SHG signal ~ 773 nm wavelength (Fig. 4 (a)), verifying the frequency doubling process inside the LN microdisk. To further confirm this, we monitored the scattered optical signal using a silicon CCD camera (SUMIX-M81M, black and white, sensitive to wavelength < 1 μm) mounted on top of the pumped LN microdisk. Inset of Fig. 4 (a) shows the camera image, indicating the frequency doubled light traveling around the microdisk perimeter. Finally, in order to investigate the power dependence of the SHG process, we replaced the CCD camera with a highly-sensitive visible detector (HAMAMATSU, H10721-01) and measured SHG power scattered vertically for different input telecom laser power levels. The results (Fig. 4 (b)) show quadratic relationship between the output and input powers, expected from the second order nonlinear process [31]. Inset of Fig. 4 (b) shows the linear fit of double-log power dependence, with a fitted slope of 1.99 ± 0.02. We calibrated the actual power received by the detector using an attenuated HeNe laser (Melles Griot, 632.8 nm) and a power meter (Newport Model 1928-C). The calculated detector sensitivity was around 40.0 A/W. Considering the detector quantum efficiencies at different wavelengths, given in its datasheet, the detector sensitivity at 773 nm was estimated to be 1.14 A/W. The in-coupled power was calculated by assuming fiber taper losses were evenly distributed on the input and output port. When 1.83 mW of input power was coupled into the resonator, the collected SHG signal was 8.34 nA, which corresponds to 7.30 nW. Assume the SHG light was distributed uniformly in all directions, 2% of the generated light could be collected from the objective lens (NA = 0.28). Indeed the SHG light is likely coupled to higher order microdisk modes (discussed next) and the emission will be stronger in the plane of the device, which means our estimation likely gives a lower bound of the generation efficiency. The internal second harmonic intensity is thus calculated to be at least 0.365 µW. From this result, we extract a conversion efficiency of $1.99 \times 10^{-4}$ (-37 dB). When normalized by in-coupled power, the device gives a conversion efficiency of 0.109 $W^{-1}$. The normalized conversion efficiency is higher than that reported in Ref. [21] since in that case SHG signal is collected from the same fiber taper and the collection efficiency is usually low for SHG wavelengths. In real applications, it is more efficient to collect SHG signal from a bus waveguide and then couple the light into a single mode fiber using either butt coupling [32] or grating coupler [13].

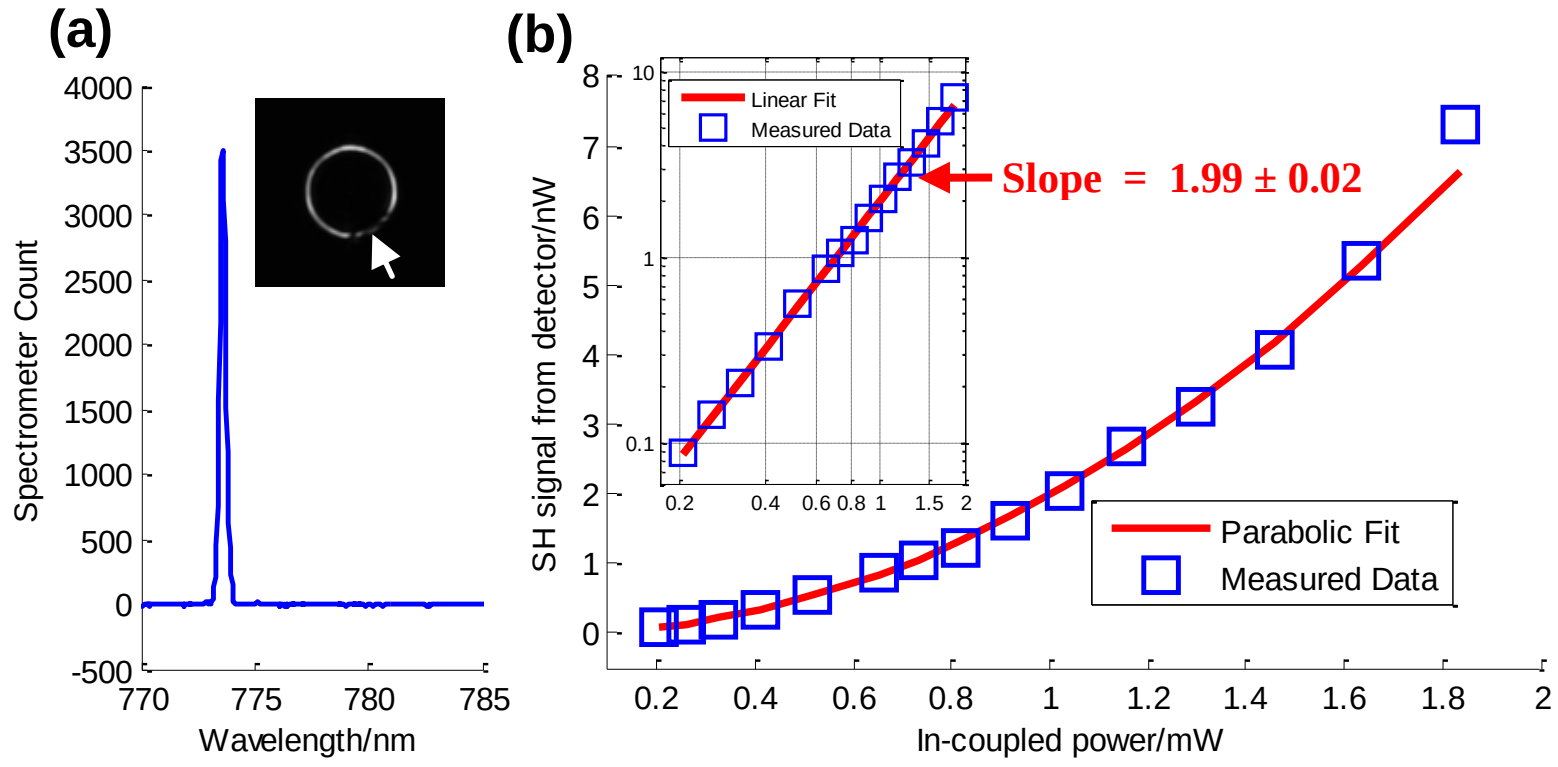

Fig. 4. Second harmonic generation. (a) Spectrometer data showing SHG peak at 773 nm when pumped at 1546 nm, indicating frequency doubling. Inset: black-and-white CCD camera image revealing scattered SHG light travelling around the LN microdisk perimeter. The fiber taper is positioned at the relatively dimmed area, indicated by the arrow. (b) Input-output power dependence and its quadratic fit. Inset shows the linear fit of double-log curve with a slope of 1.99 ± 0.02.

Owing to the wealth of optical modes available in the 770 nm wavelength range (Fig. 3 (a)), SHG could be observed in most high-Q telecom resonances. Indeed, we note that different phase matching conditions apply to different combinations of the fundamental/second harmonic polarizations, offering more than one phase matching possibilities. For the TE/TM case, since the nonlinear susceptibility is symmetric in plane ($\chi^{(2)}_{xxz} = \chi^{(2)}_{yyz}$), the usual phase matching condition applies: the azimuthal mode number of the second harmonic ($n$) equals twice that of the fundamental wavelength ($m$), i.e. $n = 2m$. For the TE/TE case, on the other hand, nonlinear susceptibility is non-zero only in one direction ($\chi^{(2)}_{yyy} \neq 0$, $\chi^{(2)}_{xxx} = 0$), giving additional phase matching conditions, i.e. $n = 2m \pm 1$, $n = 2m \pm 3$. Rigorous derivation of these arguments can be derived from the methods introduced in Ref. [33]. Our simulation (data not shown) shows that these phase-matched modes are usually closely packed near SHG wavelengths, so that there is almost always finite mode overlap between fundamental and second harmonic wavelengths. Importantly, we believe that our current devices are not perfectly mode-matched due to a limited parameter space available for dispersion engineering in microdisk resonators. A more efficient SHG could be achieved in a dispersion engineered ring resonator for example [24], which will be considered in our future work.

### *3.4 Power handling and time-domain measurements*

Due to the high quality factors in our devices, optical bistability effects [34] were observed at elevated input powers, as is shown in Fig. 5 (a). As the pump laser wavelength is tuned across the LN microdisk cavity resonance from blue to red at sufficiently high input power, increased intracavity energy results in a shift in local refractive index. The cavity stays in resonance with the laser wavelength until it abruptly jumps out of resonance when the red detuning becomes too large. Optical bistability in our devices also leads to a clear hysteresis when setting the laser at different red-detuned frequencies and monitoring input-output power relations for both upwards and downwards power sweeps (Fig. 5 (b)). A fiber coupled EO modulator (Lucent 2623NA) was installed into the input port of the fiber taper, before the inline fiber polarizer, to modulate the input power in real time. The in-coupled powers in Fig. 5 (a-b) have accounted for losses from the tapered fiber transmission (~ 5 dB) and the EO modulator (~ 5 dB), as well as an on-resonance coupling efficiency near 50% (estimated from the resonance transmission depth). The onset of the power hysteresis loop is seen to open at detuning larger than 80 pm.

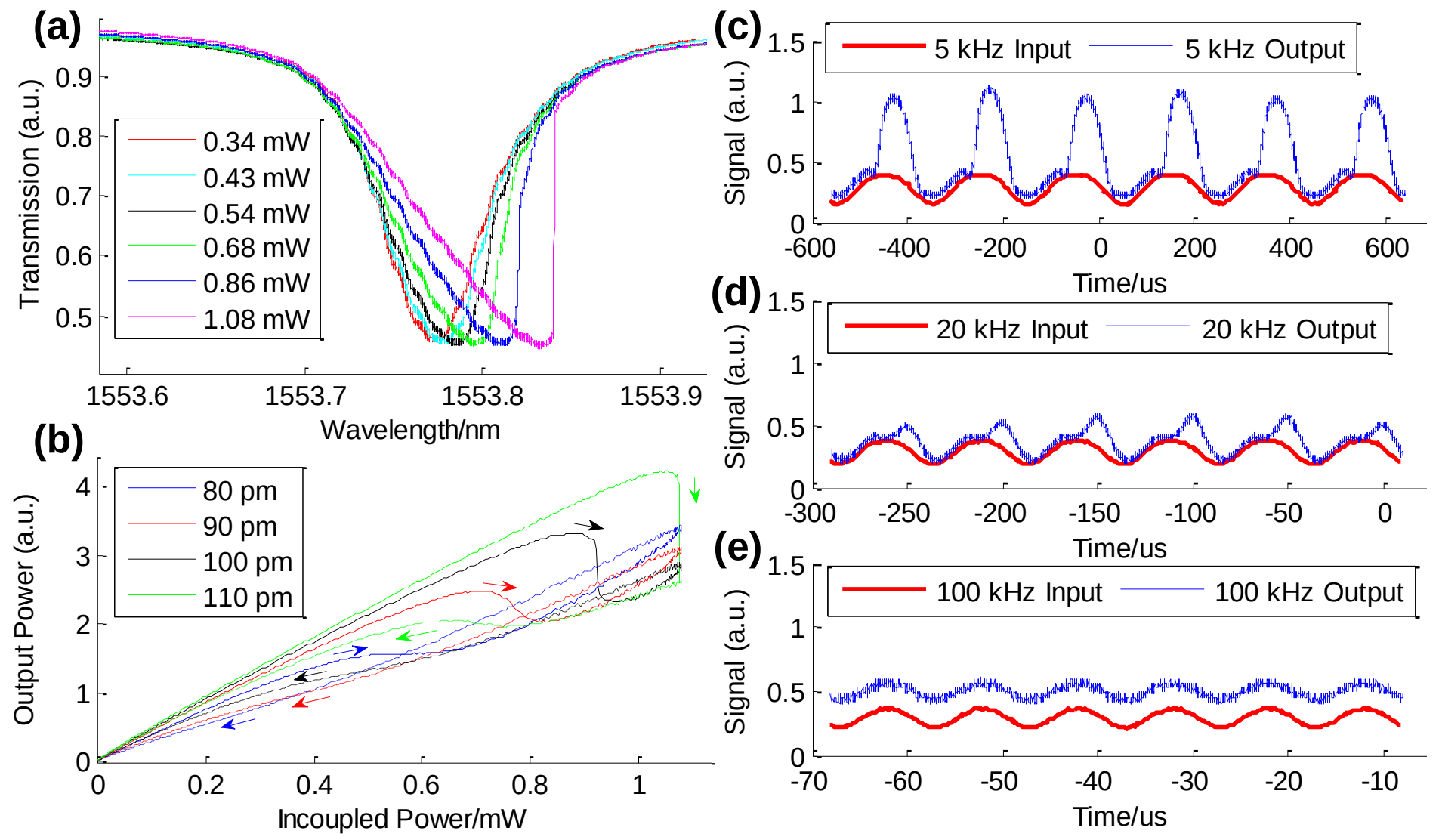

Fig. 5. Optical bistability and its time-domain response. (a) Transmission spectra of a LN microdisk resonance at different in-coupled power levels, showing red-shift of resonant frequency and optical bistability at elevated input power. The baselines of each spectrum are normalized by input power. (b) Power hysteresis curves at different red-detuned laser wavelengths (legend indicates detuning). The arrows show the directions to which the laser power is swept. (c-e) Time-domain power responses showing both input optical signals (red) and output detector signals (blue) at different modulation frequencies: (c) 5 kHz, (d) 20 kHz and (e) 100 kHz.

The observed optical bistability can originate from a number of nonlinear refractive index tuning mechanisms [35, 36], including absorption-induced thermal effects (slow) and third-order optical nonlinearities (fast). To elucidate its origin, we further investigated the bistable response through time-domain measurements (Fig. 5 (c-e)). Here, the pump laser wavelength was red-detuned from the cavity resonance and the input power is modulated with a sinusoidal signal. The output signal from a fast InGaAs photodetector (New Focus Model 1811 125-MHz photoreceiver) was monitored using an oscilloscope. For low frequency modulation (5 kHz), distortion is clearly present, and is attributed to output signal being switched from one bistable state to the other. Distortion (bistable behavior) starts to abate as the modulation frequency is increased, and completely disappears at 100 kHz. Based on these results, we estimate a time constant of the observed bi-stable behavior to be on the order of ~ 10 µs. This indicates that the origin of bistability is likely thermal, induced by absorption from defects inside LN and/or various surface states resulting from our fabrication process. These effects may be eliminated by further optimization of both our fabrication procedure and the LNOI platform itself.

## 4. Conclusions and outlook

We have demonstrated the fabrication and experimental characterization of high-Q LN microresonators operating in both visible and telecom bands by directly structuring thin film LN. Our fabrication method involves standard EBL, RIE and wet etching and can be used to realize other types of optical components in LN. The highest Q-factor we have measured ($1.02 \times 10^5$) is much higher than previous results using single-step photolithography or FIB [18, 19], and is comparable with more recent reports using much more complicated processes [20, 21]. (We note, however, that during the preparation process of this manuscript, Wang et al. reported LN microdisk resonators with Q-factors as high as 484,000 [37], realized using techniques similar to ours and geared towards optomechanics experiments. In that case the microdisks are fabricated on Si substrate, which has a higher index than LN and therefore is not suitable for integrated optics applications, e.g. waveguide and ring resonator.) Using these devices we have demonstrated on-chip SHG, with an estimated internal conversion efficiency of 0.109 $W^{-1}$.

We believe that our work will pave the way to a wide range of functional devices and systems based on thin LN film technology. One example is microresonator based EO modulator with low operating voltage. Assuming frequency tunability of 0.28 GHz/V [18] and cavity $Q \sim 5 \times 10^4$ (common in our devices), as little as 7 V of applied voltage is needed to switch the cavity on- and off-resonance. Another example is efficient nonlinear wavelength conversion: recently SHG efficiency as high as 9% at 30 μW in-coupled power has been experimentally demonstrated in high-Q mm-size LN disk resonators realized by polishing methods [38]. We have also theoretically predicted that close-to-unity SHG efficiency is possible with careful phase-matching in a doubly-resonant microring resonator with a similar size as our devices [24]. We believe that these results are well within the reach of the platform demonstrated in this work. Therefore, the fabrication methods and high-performance devices described here will serve as an important milestone in future development of integrated LN-based nonlinear nanophotonics.

**Acknowledgements**

This work was supported in part by DARPA QuINESS Program. Fabrication was performed at the Center for Nanoscale Systems (CNS) at Harvard University. M.J. Burek and H.A. Atikian were supported in part by the Natural Science and Engineering Council (NSERC) of Canada and Harvard Quantum Optics Center (HQOC). The authors thank Dr. Hui Hu from NANOLN, Dr. Ling Xie from CNS and Dr. Jonathan C. Lee for valuable discussions, and thank Johan Israelian for his assistance in fabrication.